\documentclass[]{aipproc}
\layoutstyle{6x9}
\usepackage{rotating}
\def\apj{ApJ}
\def\mnras{MNRAS}

\def\aap{A\&A}                   
\def\apjs{ApJS}                  

\def\ssr{Space Sci. Rev.}

\def\memsai{Mem. Soc. Astron. Italiana}

\def\lsim{\;\raise0.3ex\hbox{$<$\kern-0.75em\raise-1.1ex\hbox{$\sim$}}\;}
\def\gsim{\;\raise0.3ex\hbox{$>$\kern-0.75em\raise-1.1ex\hbox{$\sim$}}\;}

\def\cmc{\rm ~cm^{-3}}

\def\kms{\rm ~km~s^{-1}}

\def \kms {\rm ~km~s^{-1}}

\usepackage{color}
\newcommand{\tSedov}{t_\mathrm{Sedov}}
\newcommand{\EnSN}{E_\mathrm{SN}}
\newcommand{\Mej}{M_\mathrm{ej}}
\newcommand{\Rsh}{R_\mathrm{sh}}
\newcommand{\ush}{u_\mathrm{sh}}
\newcommand{\pmax}{p_\mathrm{max}}
\newcommand{\Emax}{E_\mathrm{max}}

\newcommand{\xx}[1]{\!\times\!10^{#1}}
\newcommand{\muG}{$\mu$G}
\newcommand{\SNRJ}{SNR RX J1713.7-3946}
\newcommand{\tSNR}{t_\mathrm{SNR}}
\newcommand{\Msun}{\mbox{$M_{\odot}\;$}}
\newcommand{\dMdt}{dM/dt}
\newcommand{\Vwind}{V_\mathrm{wind}}
\newcommand{\Twind}{T_\mathrm{wind}}
\newcommand{\SigWind}{\sigma_\mathrm{wind}}

\def\lsim{\raise0.3ex
  \hbox{$<$\kern-0.75em\raise-1.1ex\hbox{$\sim$}}\,}
\def\gsim{\raise0.3ex
  \hbox{$>$\kern-0.75em\raise-1.1ex\hbox{$\sim$}}\,}
\makeatletter
\def\@oddhead{\reset@font{\footnotesize \vbox to
0pt{\vss\vtop{\hbox{}\hbox{\bf AIP Conference Proceedings v.1505,
pp. 46-55, 2012}}}}\hfill\thepage} \makeatother
\begin{document}

\title{Galactic Cosmic Ray Origin Sites: Supernova Remnants and Superbubbles }

%
%

\classification{98.38.Mz, 98.35.-a, 98.35.Hj, 98.38.-j, 95.85.Bh, 95.80.+p}

\keywords{supernova remnants -- Milky Way}

\author{A.~M.~Bykov$^{1,3}$, D.~C.~Ellison$^2$, P.~E.~Gladilin$^{1,3}$, S.~M.~Osipov$^{1,3}$}{address={$^{1}$A.F.Ioffe Institute of Physics and Technology,
Saint-Petersburg, Russia,
\\ $^{2}$North Carolina State University, Box 8202, Raleigh, NC
27695, U.S.A.\\ $^{3}$Saint-Petersburg State Polytechnical
University, Saint-Petersburg, Russia}}

\begin{abstract}
We discuss processes in galactic cosmic ray (GCR) acceleration sites
- supernova remnants, compact associations of young massive stars, and
superbubbles. Mechanisms of efficient conversion of the mechanical
power of the outflows driven by supernova shocks and fast stellar
winds of young stars into magnetic fields and relativistic particles
are discussed. The high efficiency of particle acceleration in the
sources implies the importance of nonlinear feedback effects in a
symbiotic relationship where the magnetic turbulence required to
accelerate the CRs is created by the accelerated CRs themselves.
Non-thermal emission produced by relativistic particles (both those
confined in and those that escape from the cosmic accelerators) can be used
to constrain the basic physical models of the GCR sources. High
resolution X-ray synchrotron imaging, combined with
GeV-TeV
gamma ray spectra, is a powerful tool
to probe the maximum
energies of accelerated particles. Future MeV regime spectroscopy
will provide unique information on the composition of accelerated
particles.
\end{abstract}

\maketitle
\section{Introduction}
A hundred years after the discovery of cosmic rays (CRs) there is little
doubt that supernova remnants (SNRs) are the sources of the bulk
population of CRs in the Galaxy
with energies up to a few times 10$^{15}$ eV (see e.g., \citep{Hillas05,AhaByk12}) and may well produce CRs to
10$^{18}$  eV (e.g., \citep{pzs10}) .
Diffusive shock acceleration
(DSA) is the most likely mechanism to accelerate CR particles in the
forward and reverse shocks of the SNR shells. The high
efficiency of the DSA mechanism has been
demonstrated by nonlinear modeling (see
e.g., \citep{bell87,je91,MalDru01,AB05,VladByk08}). However, SNRs do
not comprise a uniform population of similar objects. There is a
wide variety of different SNRs since their evolution, CR acceleration,  and non-thermal emission
depend strongly on both the circumstellar and interstellar environments \citep{c05,bceu00,bv04,gac09,pzs10,mc12}.

It is important that a substantial fraction of
core-collapsed supernovae likely occur in compact clusters of young
massive stars and are associated with molecular clouds. The most
massive stars explode first and their shocks are propagating through
a complex environment created by both the very strong radiation of young
massive stars and radiatively driven stellar winds.
We discuss in \S1 the results of modeling
CR acceleration and the nonthermal emission of a supernova exploding in the stellar
wind of its progenitor star   following \citep{ellisonea12}.
In addition to an isolated SNR, in a compact rich cluster of young massive stars
the distances between the young stars may be about 10-15 pc and therefore  CRs may also be accelerated when the MHD flows produced by the supernova shell interact with a stellar wind from a nearby massive star.
%
Particle acceleration in the region where the expanding supernova
shell approaches a powerful stellar wind is discussed in \S2
\citep[see also][]{bgo11,bgo12}.
At a later stage of the young stellar cluster evolution
(an age of about ten million years) multiple
supernova explosions with great energy release in the form of shock
waves inside a  superbubble is a favorable site for  particle acceleration.
The collective acceleration from both
stellar
winds of massive stars and core collapsed supernovae
in superbubbles was discussed in
\citep[][]{bykov01,bt01,fm10,lingen12}.

\section{CR acceleration by core-collapsed SNRs}
Early type stars with masses above $\sim 16\,\Msun$
(of B0~V type and earlier) are thought to create hot,
low-density
bubbles  with radii $\sim 10$\,pc surrounded by a
massive shell of matter swept up from the parent cloud by the
stellar wind.
A SN exploding in the early
stages (i.e., the first few million years) of the OB association evolution would interact with a
cavern created by the pre-supernova wind.
In this case, a strong supernova shock propagates for a few thousand
years in tenuous circumstellar matter with a velocity above
$10^3$\,km/s before reaching the dense massive shell.
Magnetic field
fluctuations in the shock vicinity may be highly amplified by
instabilities driven by the CR-current and CR-pressure
gradient in the strong shocks \citep[][]{bell04,boe11,schureea12}.
This is an important factor for determining the  highest energy
particles accelerated by the shocks.

\begin{figure} \label{Fig:SNRXJ}
{\includegraphics[scale=0.65]{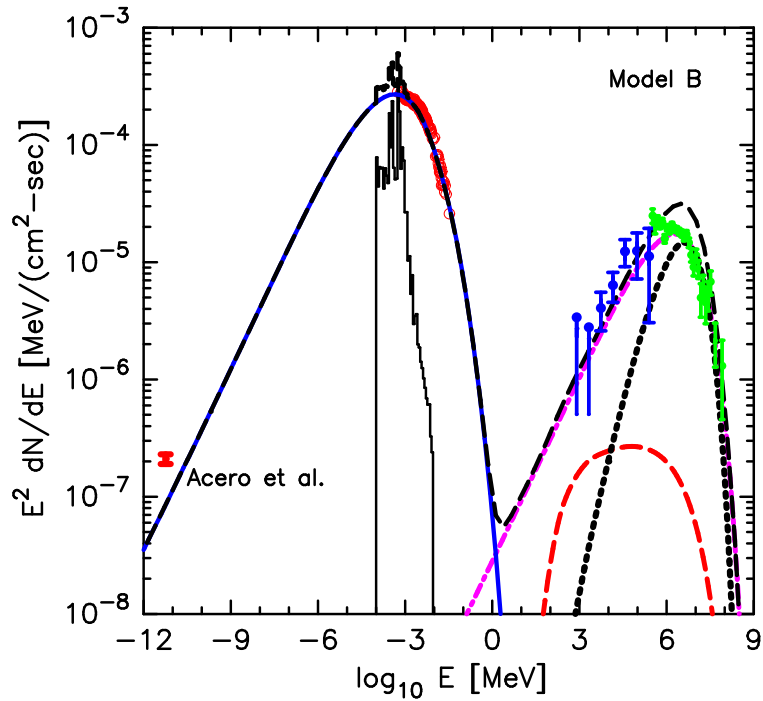}
\caption{A model fit to SNR RXJ1713 observations from \citep{ellisonea12}. The different
emission
processes are: synchrotron (solid  curve), IC (dot-dashed curve),
pion from trapped CRs (dashed red curve), pion from escaping
CRs (dotted  curve), and thermal X-rays (solid curve with X-ray line features). The
dashed black curve is the summed emission.}}
\end{figure}

A nonlinear, spherically symmetric  model of the core-collapsed
SNR RX J1713.7-3946 that includes a hydrodynamic simulation of the remnant
evolution coupled to the efficient production of CRs by
DSA at the supernova forward shock was studied in \citep{ellisonea12} (see Figure~\ref{Fig:SNRXJ}). High-energy CRs that escape from the
forward shock  are propagated in surrounding dense material that
simulates either a swept-up,
pre-supernova shell or a nearby molecular cloud. The continuum
emission from trapped and escaping CRs, along with the thermal X-ray
emission from the shock-heated ISM behind the forward shock, integrated over
the remnant, was compared against broadband observations.
Overall, the GeV-TeV emission is dominated by inverse-Compton (IC)
emission from CR
electrons if the supernova is isolated regardless of its type, i.e.,
not interacting with a $\gg\! 100\,\Msun$ shell or cloud.
If the SNR is interacting with a much larger mass
$\gsim 10^4\,\Msun$, pion production by the escaping CRs may dominate the TeV
emission, although a precise
fit  at high gamma-ray energies will depend on the still uncertain details of how
the highest energy CRs are accelerated by, and escape from, the forward shock (FS).
Importantly, even though CR electrons
dominate the
GeV-TeV emission, much more energy is put into CR ions and the efficient production of CR  ions
 is an essential part of our leptonic model.

An important factor that allows a good fit to the broadband
spectrum of \SNRJ\ with leptons, particularly including the highest energy HESS points, stems from the fact that the pre-SN wind magnetic field, in which the SN explodes, is
considerably
lower than 3\,$\mu$G due to the expansion of the wind. This allows the
electrons to be accelerated to higher
energies before radiation losses dominate.
We take a pre-SN wind with speed $\Vwind$,
mass-loss rate $\dMdt$, and temperature $\Twind$, all of which are
assumed constant. The parameter $\SigWind$  determines the wind magnetic field at a
radius $R$ from the explosion according
to
\begin{equation}\label{eq:sigma}
B_0(R) = \frac{(\SigWind \Vwind \dMdt)^{1/2}}{R}
\ .
\end{equation}
The constant parameter $\SigWind$ is the ratio of magnetic field
energy density
to kinetic energy density in the wind and can be related to
properties of the star by \citep[e.g.,][]{cl94,walderea11}
\begin{equation}\label{eq:sigmaTwo}
\SigWind \propto \frac{B_{\ast}^2 R_{\ast}^2 }{(\dMdt)v_w}\times
\left(\frac{v_r}{v_w}\right)^2
\ .
\end{equation}
Here, $B_{\ast}$ denotes the surface magnetic field of the star,
$R_{\ast}$ is the stellar radius, $v_r$ is the star rotation velocity,
and $v_w=\Vwind$ is the terminal speed of the wind.
Obtaining $\SigWind \sim 0.03$ by fitting the
spectrum of SNR RX J1713.7-3946, as was done in \citep{ellisonea12}, constrains the progenitor star
parameters.
As noted by \citep{walderea11}, values of $\SigWind \ll 1$ indicate
that the stellar wind dominates
the magnetic field, producing a roughly radial field far from the
star.

\begin{figure} \label{fig:prof}
\includegraphics[scale=0.65]{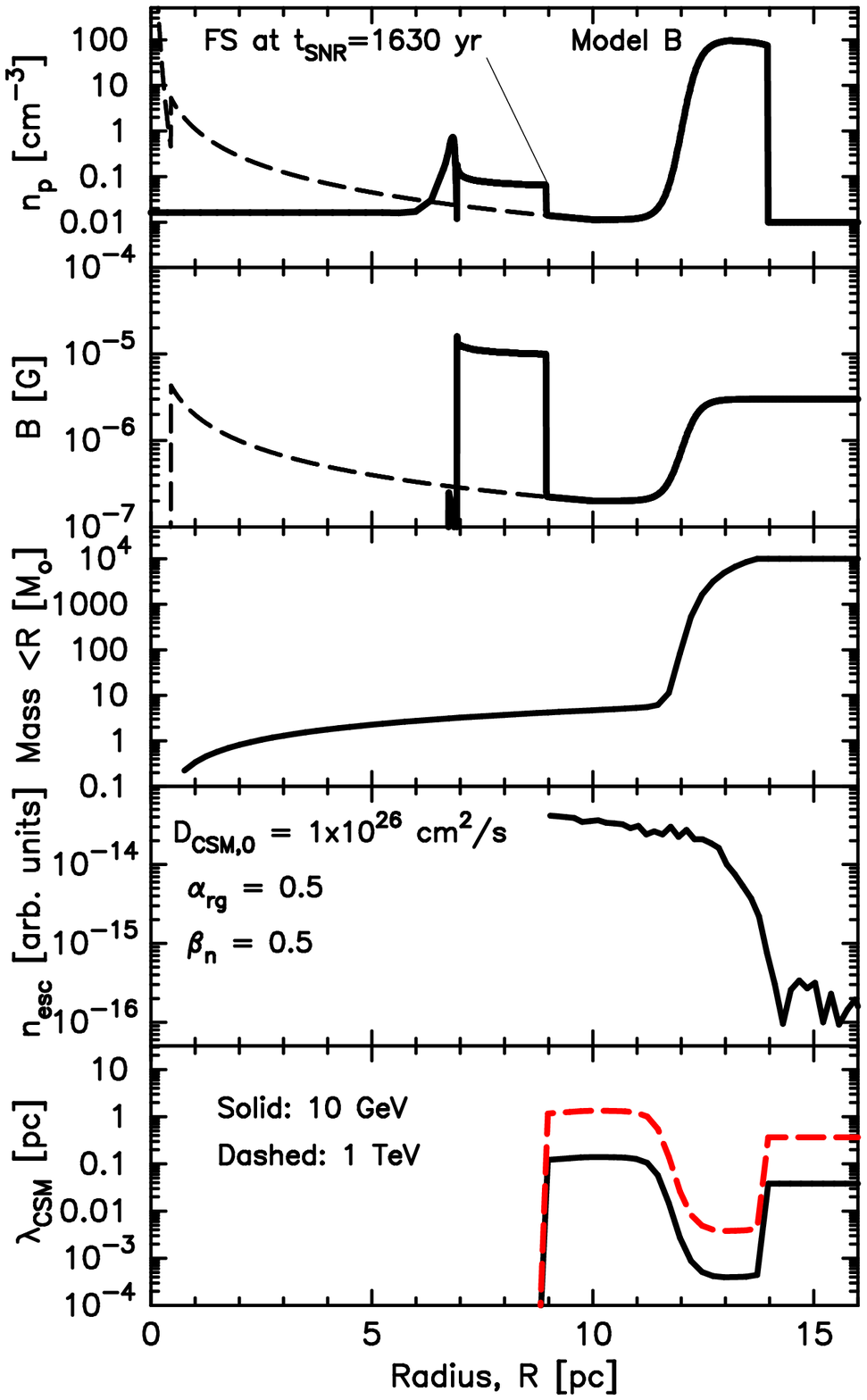}
\caption{Radial profiles for \SNRJ\ from \citep{ellisonea12}. The top two panels show the proton number density and the
magnetic field as a function of distance from the center of the SNR. In each of these two panels, the dashed curve is the
profile at the beginning of the simulation and the solid curve is the
profile at $\tSNR=1630$\,yr. The third panel shows the mass within $R$
at $t=0$ and the fourth panel shows the escaping CR number density.
The diffusion parameters used in \citep{ellisonea12} are listed in
the fifth panel. Escaping CRs are only followed beyond the FS and
they leave the spherically symmetric simulation freely at the outer
radius of  $\sim 16$\,pc. }
\end{figure}

For the parameters used in \citep{ellisonea12}
\citep[see also][]{EllLee12}, the upstream magnetic field at the forward shock at the current age of \SNRJ\ is $\sim 2\xx{-7}$\,G and
the $\sim 10$\,\muG\ field immediately downstream from the shock is amplified by a factor of about 50 above the unshocked wind magnetic field (see Figure~\ref{fig:prof}).
The low unshocked wind field reconciles efficient CR acceleration, and the accompanying strong magnetic field amplification (MFA), with the low magnetic field that is required for leptons to produce the highest energy gamma-ray emission with IC.
It may  also
provide the relatively wide ($\sim$ parsec-scale size) profile of the
IC emission consistent with the observed  TeV profile.  The discussion of the possible effect of the reverse shock can be found in
\citep{za10}.
We confirm that the circumstellar medium (CSM) can  strongly influence the non-thermal
emission from  core-collapsed   SN Ib/c and  SNIIb SNRs, and emphasize again that the magnetic field of
%
the  progenitor massive star may be well below
the average interstellar medium (ISM) values of a few micro-Gauss alleviating an apparent contradiction between the low postshock
magnetic field values required  by the leptonic origin of high energy gamma-rays with the
strong magnetic field amplification expected in efficient DSA.

It is also important to emphasize that, even though electrons dominate the high-energy emission, the DSA model we are discussing accelerates CR
ions efficiently. The results we illustrated here
(Figures~\ref{Fig:SNRXJ} and \ref{fig:prof}) place 25 to 50\%
of the forward shock ram
kinetic energy flux into relativistic ions at any
instant. Only 0.25\% or less of the
instantaneous ram kinetic energy
flux goes into relativistic electrons. Leptons dominate the model emission
simply because leptons radiate far more efficiently than ions, not
because ions are missing. Furthermore, the best-fit parameters for
this remnant result in maximum
proton energies of $\sim 10^{14}$\,eV. Iron nuclei would be
accelerated to $\sim 26\times 10^{14}$\,eV, well into the CR ``knee'' regime.

\section{CR acceleration by colliding SNR-wind shocks}
\begin{figure} \label{Fig:SNR_SW}
\resizebox{16cm}{9cm}{
\includegraphics{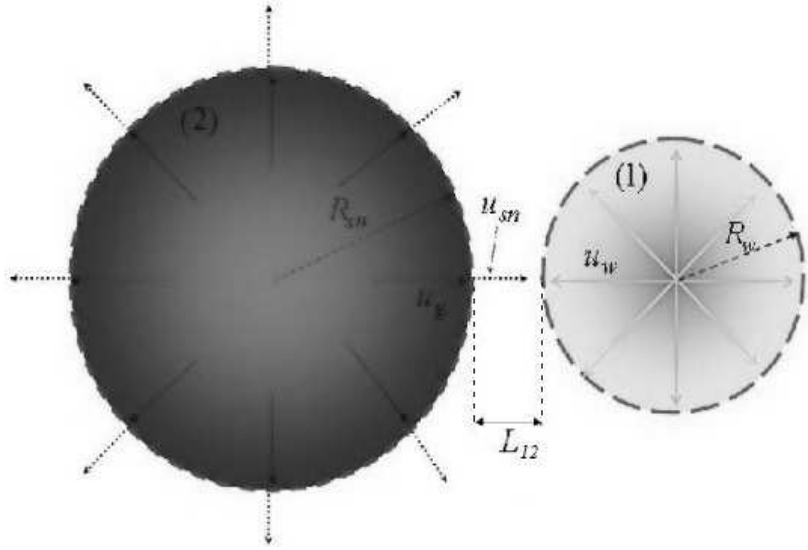}}
\caption{Simplified SNR--stellar wind system as described in the text. The supernova
remnant
 is on the left and the stellar wind termination shock is on the right.}
\end{figure}

The blast wave
of the SNR is expected to accelerate the wind material producing
ultra-relativistic ions and electrons. The highest energy CRs
escape from the SNR forward shock and can reach the termination
shock of the stellar wind of a nearby massive star. In
Figure~\ref{Fig:SNR_SW} we show a highly simplified situation where
the SNR shock approaches the
the stellar
wind from a young OB-star. For efficient acceleration in such
a system, the distance $L_{12}$ between the shocks should be about a pc and
we estimate that this stage may last about
1,000 yrs.

We modeled the
energetic particle acceleration in the region where the expanding
supernova shell is approaching a powerful stellar wind of a young
massive star as it is illustrated in Figure~\ref{Fig:SNR_SW}.
At the
evolutionary stage where the mean free path of the highest energy CR is
comparable to the distance between the two shocks, $L_{12}$, the system is
characterized by a very hard spectrum of accelerated CRs as discussed in \citep{bgo11}. The hard CR spectrum produces  unusual spectral energy distributions (SEDs) of synchrotron and IC emission as
illustrated in Figure~\ref{Fig:Spectrum}.
The SED is derived for the moment of close approach of the shell
with the stellar wind with $L_{12} \lsim 0.1$ pc.  For comparison, we show in Figure~\ref{Fig:Spectrum}, with the dotted lines, the SED
produced by an isolated SNR shock at a similar
age.

\subsection{The maximum CR energy}
The effect of different environments in which an isolated SNR may evolve on the maximum CR energy, and the time evolution of the maximum energy, was discussed in  \citep{dtp12}.
In their model, the maximum CR energy is generally reached in the ejecta-dominated stage, much before the start of the Sedov-Taylor stage. It was concluded that, for SNe evolving within the winds of their massive stars,
the maximum energy is reached very early in the evolution.

As shown in \citep{bgo12}, a supernova exploding in a young stellar
cluster and interacting with a nearby strong stellar wind will
produce higher maximum energy CRs  in the colliding MHD flows
compared to an isolated SNR. Furthermore, the SNR can produce
high-energy CRs at the Sedov-Taylor evolution stage.

The evolution of a SNR, after the star explosion, is characterized (in
chronological order) by the free expansion phase, the Sedov-Taylor stage,
and the radiative stage. During the free expansion phase the forward shock
velocity of the ejecta is approximately constant. At the Sedov-Taylor stage,
the ejecta starts to decelerate when the swept-up mass becomes
comparable to the ejecta mass. The transition from the free
expansion phase to the Sedov-Taylor stage occurs around the time $\tSedov$
(see e.g., \citep{helderea12}) where

\begin{equation}
\tSedov =2.6 \left(\frac{\EnSN}{10 ^{51} \mathrm{erg}}\right)^{-1/2}
\left(\frac{\Mej}{8 M_{\odot}}\right)^{5/6} \left(\frac{n_0}{0.1
~\cmc }\right)^{-1/3} \mathrm{kyr}. \label{eq:tsed}
\end{equation}
Here,  $n_0$ is the number density of the interstellar medium,
$\EnSN$ is the SNR explosion energy, $\Mej$ is the mass of the ejecta, and
$M_{\odot}$ is the mass of the Sun.

The evolution of the forward shock radius $\Rsh(t)$ and shock
velocity $\ush(t)=d\Rsh/dt$ during the Sedov phase of the SNR
expansion are given by the following equations:

\begin{equation}
\Rsh =25.5  \left(\frac{\Mej}{8 M_{\odot}}\right)^{1/3}
\left(\frac{n_0}{0.1 ~\cmc }\right)^{-1/5}
\left(\frac{t}{\tSedov}\right)^{2/5} \mathrm{pc},
\end{equation}
\begin{equation}
\ush =3.7 \cdot10^3
\left(\frac{\EnSN}{10^{51} \mathrm{erg}}\right)^{1/5}\left(\frac{n_0}{0.1
~\cmc }\right)^{-1/5}\left(\frac{t}{\tSedov}\right)^{-3/5} ~\kms.
\end{equation}

For the calculation of the maximum momentum evolution we use the
approach proposed in \citep{pz03}:
\begin{equation}
\frac{\pmax}{m_pc}=\frac{3 \chi \ush(t) \Rsh(t)}{v r_{g0}},
\end{equation}
where $r_{g0}=mc^2/eB_0$, $v\approx c$ is the particle velocity,
$B_0=1$\,\muG\ is the unperturbed magnetic field strength,
$\chi=0.04$, and
the mass ejected with the SNR explosion
is $\Mej =8 M_{\odot}$.

\begin{figure}
\resizebox{10cm}{10cm}{
\includegraphics{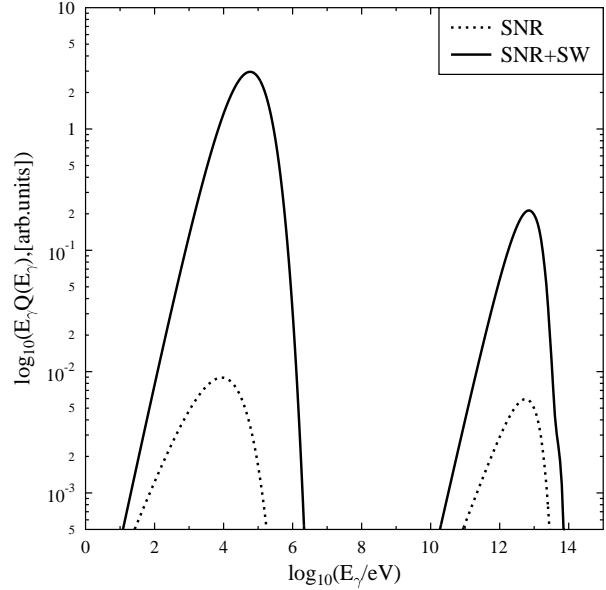}}
\caption{Spectral energy distributions  of the synchrotron
and IC emission from electrons accelerated at the
colliding shock system (solid line) compared with the emission from electrons accelerated by a single isolated shock of the same
speed (dotted line). The SEDs are given in arbitrary units. }
\label{Fig:Spectrum}
\end{figure}

\begin{figure}
\resizebox{12cm}{12cm}{
\includegraphics{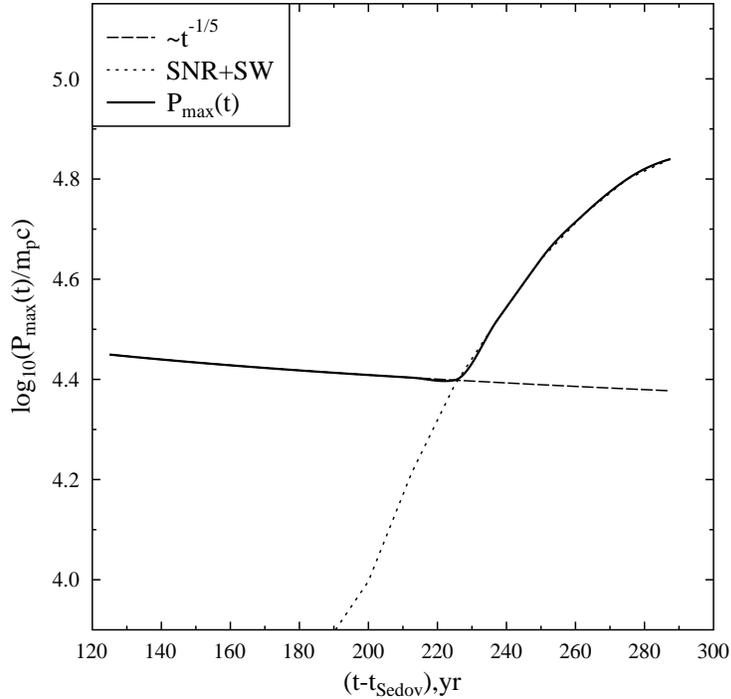}}
\caption{Maximum momentum of the accelerated particles in case of SNR-SW collision at the Sedov-Taylor phase of a SNR.
At $t-\tSedov \approx 190$\,yr, particles start to be accelerated by the two-shock, SNR--stellar wind system.
For this plot, $\tSedov$ is calculated from Eq.(~\ref{eq:tsed}).
The dashed curve is $\pmax$
for the case of ordinary SNR evolution, the dotted curve is $\pmax$ for the
two-shock system when the SNR blast wave  starts to ``feel'' the stellar
wind of the nearest star, and the solid line is the
resulting $\pmax$. }
\label{Fig:Sedov}
\end{figure}

In Figure~\ref{Fig:Sedov} we illustrate the evolution of $\pmax$ accelerated in the system in which the forward
shock of the SNR approaches the strong stellar wind from a young OB star being at the Sedov-Taylor stage. The SNR-stellar wind collision may also occur
well before the Sedov-Taylor phase depending on the compactness of the stellar cluster and inter-cluster matter distribution.  At some moment during the Sedov-Taylor stage of the SNR
evolution the forward shock of the SNR is close enough to the
termination shock of the stellar wind to
produce efficient acceleration with the sharp increase in $\pmax$ shown in the figure.

The spectra of CRs accelerated in the colliding flows are harder than
those from DSA produced by a single shock \citep{bgo11}.
In the Figure~\ref{Fig:Spectrum} we compare the
IC
and synchrotron emission from the electrons accelerated in a
colliding shock with that from electrons from a single shock.
For this example, the local interstellar photon spectrum was
constructed taking into account background IR fluxes typical for
OB-associations, as measured by \citep{Saken92}. The electron
distribution functions used in the spectra calculations were computed
with a time-dependent model that accounts for nonlinear flow
modifications in the vicinity of the shocks, and radiation
losses for the electrons. It is clearly seen that
the emissivity of the SNR-SW
system is considerably higher than
that for the single SNR shock due to the
hard electron spectrum up to $10^4$ GeV.



\section{Conclusions}
Supernova remnants are the most likely candidates to accelerate galactic CRs up to the knee regime and
perhaps beyond the knee. Core-collapsed supernovae are thought to be statistically the dominating class of supernova events.
The effect of the clustering of core-collapsed supernova progenitors, and the interaction of supernova ejecta with the circumstellar
 matter  produced by the wind of the progenitor star, affect both
the maximum energy CRs can obtain and the CR composition.
In this paper we discussed briefly a few recent results illustrating the effects of both the strong wind of the progenitor massive star, and the collision of a supernova shock with the
powerful wind of a nearby young massive star. In particular, we emphasized the importance of the
 structure of the magnetic field in the stellar wind of the progenitor star. In the case of slow rotation of the progenitor star,
the wind magnetic field at large distances from the star may be well below the typical $\sim 3$\,\muG\ value for the ISM. In this case, the strong SNR forward shock
can accelerate CR ions efficiently with the expected strong MFA
and still have low enough fields behind the shock
(i.e., $\sim 10$\,\muG) so relativistic electrons can obtain high enough energies
to produce the high-energy gamma-ray emission via IC before radiation losses take over.

In the case of a supernova explosion in a compact cluster of young massive stars, the collision of the supernova shock with the stellar wind of a nearby
star is likely and the combined effect of shock and stellar wind
may result in a substantial increase
in the maximum CR energy, $\Emax$.
This increase in $\Emax$ can occur
even during the stage when the maximum CR energy from the isolated SNR was declining (see Figure~\ref{Fig:Sedov}).
The characteristic spectrum of non-thermal emission from the SNR-SW collision is expected to be very hard (see Figure~\ref{Fig:Spectrum}), a prediction that can be tested observationally.

\begin{theacknowledgments}
A.M.B. and D.C.E. thank F.~Aharonian, W.~Hofmann and F.~Rieger for
the excellent Symposium ``High Energy Gamma-Ray Astronomy''. The
work was supported in part by
 the Russian government grant
11.G34.31.0001 to the Saint-Petersburg State Polytechnical
University, and also by the RAS Programs (P21 and OFN 16), by the
RFBR grant 11-02-12082-ofi-m-2011, by Ministry of Education and
Science of Russian Federation (Agreement No.8409, 2012). The
numerical simulations were performed at JSCC RAS and the SC at Ioffe
Institute. D.C.E. acknowledges support from NASA Grant NNX11AE03G.
\end{theacknowledgments}

\bibliographystyle{aipprocl}

\begin{thebibliography}{10}
\providecommand{\enquote}[1]{``#1''} \expandafter\ifx\csname
url\endcsname\relax
  \def\url#1{\texttt{#1}}\fi
\expandafter\ifx\csname urlprefix\endcsname\relax\def\urlprefix{URL
}\fi

\bibitem{Hillas05}
A.~M. {Hillas}, \emph{Journal of Physics G Nuclear Physics}
\textbf{31}, 95--+
  (2005).

\bibitem{AhaByk12}
F.~{Aharonian}, A.~{Bykov}, E.~{Parizot}, V.~{Ptuskin}, and
A.~{Watson},
  \emph{\ssr} \textbf{166}, 97--132 (2012).

\bibitem{pzs10}
V.~{Ptuskin}, V.~{Zirakashvili}, and E.-S. {Seo}, \emph{\apj}
\textbf{718},
  31--36 (2010).

\bibitem{bell87}
A.~R. {Bell}, \emph{\mnras} \textbf{225}, 615--626 (1987).

\bibitem{je91}
F.~C. {Jones}, and D.~C. {Ellison}, \emph{Space Science Reviews}
\textbf{58},
  259--346 (1991).

\bibitem{MalDru01}
M.~A. {Malkov}, and L.~{O'C Drury}, \emph{Reports on Progress in
Physics}
  \textbf{64}, 429--481 (2001).

\bibitem{AB05}
E.~{Amato}, and P.~{Blasi}, \emph{MNRAS} \textbf{364}, L76--L80
(2005).

\bibitem{VladByk08}
A.~E. {Vladimirov}, A.~M. {Bykov}, and D.~C. {Ellison}, \emph{\apj}
  \textbf{688}, 1084--1101 (2008).

\bibitem{c05}
R.~A. {Chevalier}, \emph{\apj} \textbf{619}, 839--855 (2005).

\bibitem{bceu00}
A.~M. {Bykov}, R.~A. {Chevalier}, D.~C. {Ellison}, and Y.~A.
{Uvarov},
  \emph{\apj} \textbf{538}, 203--216 (2000).

\bibitem{bv04}
E.~G. {Berezhko}, and H.~J. {V{\"o}lk}, \emph{\apj} \textbf{611},
12--19
  (2004).

\bibitem{gac09}
S.~{Gabici}, F.~A. {Aharonian}, and S.~{Casanova}, \emph{\mnras}
\textbf{396},
  1629--1639 (2009).

\bibitem{mc12}
G.~{Morlino}, and D.~{Caprioli}, \emph{\aap} \textbf{538}, A81
(2012).

\bibitem{ellisonea12}
D.~C. {Ellison}, P.~{Slane}, D.~J. {Patnaude}, and A.~M. {Bykov},
\emph{\apj}
  \textbf{744}, 39 (2012).

\bibitem{bgo11}
A.~M. {Bykov}, P.~E. {Gladilin}, and S.~M. {Osipov}, \emph{\memsai}
  \textbf{82}, 800 (2011).

\bibitem{bgo12}
A.~M. {Bykov}, P.~E. {Gladilin}, and S.~M. {Osipov}, \emph{MNRAS\,
in press,\, arXiv:1212.1556}
  \textbf{} (2012).


\bibitem{bykov01}
A.~M. {Bykov}, \emph{\ssr} \textbf{99}, 317--326 (2001).

\bibitem{bt01}
A.~M. {Bykov}, and I.~N. {Toptygin}, \emph{Astronomy Letters}
\textbf{27},
  625--633 (2001).

\bibitem{fm10}
G.~{Ferrand}, and A.~{Marcowith}, \emph{\aap} \textbf{510}, A101
(2010).

\bibitem{lingen12}
R.~E. {Lingenfelter}, \emph{arXiv:1209.5728}  (2012).

\bibitem{bell04}
A.~R. {Bell}, \emph{MNRAS} \textbf{353}, 550--558 (2004).

\bibitem{boe11}
A.~M. {Bykov}, S.~M. {Osipov}, and D.~C. {Ellison}, \emph{\mnras}
\textbf{410},
  39--52 (2011).

\bibitem{schureea12}
K.~M. {Schure}, A.~R. {Bell}, L.~{O'C Drury}, and A.~M. {Bykov},
\emph{\ssr}
  \textbf{173}, 491--519 (2012).

\bibitem{cl94}
R.~A. {Chevalier}, and D.~{Luo}, \emph{\apj} \textbf{421}, 225--235
(1994).

\bibitem{walderea11}
R.~{Walder}, D.~{Folini}, and G.~{Meynet}, \emph{\ssr} \textbf{166},
145--185
  (2012).

\bibitem{EllLee12}
S.-H. {Lee}, D.~C. {Ellison}, and S.~{Nagataki}, \emph{apj}
\textbf{750}, 156
  (2012).

\bibitem{za10}
V.~N. {Zirakashvili}, and F.~A. {Aharonian}, \emph{\apj}
\textbf{708}, 965--980
  (2010).

\bibitem{dtp12}
V.~V. {Dwarkadas}, I.~{Telezhinsky}, and M.~{Pohl}, \emph{ArXiv
e-prints}
  (2012).


\bibitem{helderea12}
E.~A. {Helder}, J.~{Vink}, A.~M. {Bykov}, Y.~{Ohira}, J.~C.
{Raymond}, and
  R.~{Terrier}, \emph{\ssr} \textbf{173}, 369--431 (2012).

\bibitem{pz03}
V.~S. {Ptuskin}, and V.~N. {Zirakashvili}, \emph{\aap} \textbf{403},
1--10
  (2003).

\bibitem{Saken92}
J.~M. {Saken}, R.~A. {Fesen}, and J.~M. {Shull}, \emph{\apjs}
\textbf{81},
  715--745 (1992).

\end{thebibliography}


\end{document}